\documentclass[11pt,prd,reprint,notitlepage,nofootinbib]{revtex4-1}

\usepackage{amsmath,bm}
\usepackage{amsfonts}
\usepackage{graphicx, adjustbox}
\usepackage{aas_macros}
\usepackage{listings}
\usepackage{hyperref}
\usepackage{gensymb}

\usepackage[usenames]{xcolor}
\definecolor{mpl_blue}{HTML}{1F77B4}
\definecolor{mpl_orange}{HTML}{FF7F0E}
\definecolor{mpl_green}{HTML}{2CA02C}
\definecolor{mpl_red}{HTML}{D62728}

\usepackage{orcidlink}

\begin{document}

\title{An efficient pipeline for joint gravitational wave searches from individual binaries and a gravitational wave background with Hamiltonian sampling}

\author{Gabriel E.~Freedman\orcidlink{0000-0001-7624-4616}}
\affiliation{Center for Gravitation, Cosmology and Astrophysics, University of Wisconsin--Milwaukee, P.O. Box 413, Milwaukee, Wisconsin 53201, USA}

\author{Sarah J.~Vigeland\orcidlink{0000-0003-4700-9072}}
\affiliation{Center for Gravitation, Cosmology and Astrophysics, University of Wisconsin--Milwaukee, P.O. Box 413, Milwaukee, Wisconsin 53201, USA}

\date{\today}

\begin{abstract}
The pulsar timing array community has recently reported the first evidence of a low-frequency stochastic gravitational wave background. With longer observational timespans we expect to be able to resolve individual gravitational wave sources in our data alongside the background signal. The statistical modeling and Bayesian searches for such individual signals is a computationally taxing task that is the focus of many different avenues of methods development. We present a pipeline for performing efficient joint searches for gravitational waves originating from individual supermassive black hole binaries as well as a gravitational wave background using a Hamiltonian Monte Carlo sampling scheme. Hamiltonian sampling proposes samples based on the gradients of the model likelihood, and can both converge faster to more complicated and high-dimensional distributions as well as efficiently explore highly covariant parameter spaces such as the joint gravitational wave background and individual binary model. We show the effectiveness of our scheme by demonstrating accurate parameter estimation for simulated datasets containing low- (6 nHz) or high- (60 nHz) frequency binary sources. Additionally we show that our method is capable at more efficiently generating skymaps for individual binary sources -- maps displaying the upper limits on the gravitational wave strain of the source, $h_{0}$, as a function of sky location -- by sampling over larger portions of the full sky. Comparing against results for the NANOGrav 12.5-year dataset, we find similar reconstructed upper limits on the gravitational wave strain while simultaneously reducing the number of required analyses from 72 independent binned searches down to a single run.
\end{abstract}

\maketitle

\section{Introduction}

The first evidence of a low-frequency stochastic gravitational wave (GW) signal reported \cite{NG15yrGWB, EPTA_GWB, PPTA_GWB, CPTA_GWB} by the North American Nanohertz Observatory for Gravitational Waves (NANOGrav) \cite{NANOGrav_paper}, European Pulsar Timing Array \cite{EPTA_paper}, Indian Pulsar Timing Array \cite{InPTA_DR1}, Parkes Pulsar Timing Array \cite{PPTA_paper}, and Chinese Pulsar Timing Array \cite{CPTA_GWB} has opened a new chapter in the field of GW astrophysics. Pulsar timing array (PTA) \cite{1978SvA....22...36S, 1979ApJ...234.1100D, 1990ApJ...361..300F} collaborations search for nHz frequency GWs by analyzing the the times-of-arrival (TOAs) of radio pulses emitted by millisecond pulsars. By regularly observing such pulsars over a decades-long timespan PTAs can reach the sensitivity necessary to probe the nHz band. The recently identified stochastic GW signal displayed, to varying levels of significance, the expected Hellings-Downs (HD) \cite{1983HellingsDowns} spatial correlation signature between pulsars that is indicative of the signal being a gravitational wave background (GWB).

One possible source describing the nHz GWB is the collective signal from the population of supermassive black hole binaries (SMBHBs) present in the observable universe \cite{sesanaSMBHB2005}. All massive galaxies hold a supermassive black hole, typically of mass $10^{6} - 10^{10} M_{\odot}$, at their centers \cite{kormendyho2013}. Galactic merger events consequently lead to the formation of SMBHB systems. When the component black holes reach inspiral phase, the emission of GWs becomes the dominant force behind the system's evolution. To date there have been no confirmed observations of SMBHBs at sub-parsec separations. With the discovery of a GWB signal, a next major step for PTA science is to search for particularly loud individual binaries that may be detected amongst the stochastic ensemble within the next decade \cite{2015MNRAS.451.2417R, 2017NatAs...1..886M, 2018MNRAS.477..964K, 2022ApJ...941..119B}. Measurements of GWs from individual sources, colloquially referred to as continuous waves (CWs) due to their minimal frequency evolution, would provide useful constraints on the astrophysical environments of SMBHBs \cite{quinlanSMBHB, haimankocsisSMBHB} and could be coupled with electromagnetic observations to study galactic evolution and further multimessenger astrophysical research (eg. \cite{charisiSMBHB_EM}).

Single source searches prove inherently more computationally taxing than a comparable GWB analysis. Modeling GWs from an individual SMBHB adds to the already large PTA parameter space, and such parameters bring covariances amongst themselves as well as with other red-noise processes present in the data. The computational cost additionally compounds with increased data volume, which includes longer observation span and new pulsars added to the array. This complication is particularly apparent for efforts at combining datasets from multiple PTAs, yielding highly sensitive yet computationally overwhelming data products. Multiple techniques at exploring the CW parameter space have been developed \cite{2010arXiv1008.1782C, 2011MNRAS.414.3251L, 2013CQGra..30v4004E, 2014PhRvD..90j4028T, QuickCW}, and recent improvements have led to a 100-fold speed up of the full analysis through the use of a tailored likelihood calculation \cite{QuickCW}.

In this paper we detail an additional procedure for achieving efficient CW searches through a Monte Carlo routine established through sample proposals based in the gradient of the model likelihood. This utilizes a Hamiltonian Monte Carlo (HMC) \cite{1987HMCDuane, 2011HMCNeal} sampler to replace the random-walk nature of traditional MCMC methods with a simulation of Hamiltonian dynamics on the target probability distribution. The algorithm concentrates on drawing subsequent samples at much further distances in the multidimensional parameter space, trading pure speedups of the likelihood calculation for higher sample acceptance rates and an efficient exploration of the distribution. Within the realm of PTA science, HMC was first utilized in the development of a model-independent approach to Bayesian inference with PTA data \cite{2013PhRvD..87j4021L}, and soon after to the task of outlier removal in single-pulsar data. In a previous paper \cite{HMCandPTAs_Freedman}, we demonstrated the effectiveness of using HMC to perform Bayesian GWB searches with the full marginalized PTA likelihood. Here we extend the methods and previous results to allow for joint inference of individual binary sources and common-process signals.

This paper is outlined as follows. In Sec. \ref{sec:methods} we review the signal model of a single binary and describe the current Bayesian formalism for searching for such sources in the context of PTA data. We describe the Hamiltonian Monte Carlo sampling procedure and introduce a new pipeline, predicated on this algorithm, for performing the analyses in a more efficient manner. In Sec. \ref{sec:sim_data} we validate this pipeline against a suite of simulated PTA datasets. We assess the efficiency of this new analysis prescription on the NANOGrav 12.5-year dataset in Sec. \ref{sec:real_data}. Lastly, in Sec. \ref{sec:conclusions} we summarize and discuss opportunities for future development of this work.

\section{Methodology and Software}
\label{sec:methods}

Here we provide a brief overview of the data, PTA signal model, and likelihood function used in this paper, as well as characterize the GW signal for an individual binary. We then describe the HMC algorithm, and introduce our code and pipeline tailored to applying this method to CW searches.

\subsection{PTA Likelihood}
\label{sec:pta_signal}

First we discuss pulsar timing data and the structure of the PTA likelihood. Pulsar observational data exists in the form of pulse times-of-arrival (TOAs). After subtracting from each pulsar's TOAs a timing model comprised of parameters such as proper motion, parallax, spin period, spin period derivative, and other orbital parameters, we are left with timing residuals $\delta\mathbf{t}$ that we can characterize as a linear combination of noise sources and GW signals:

\begin{equation}
    \delta\mathbf{t} = M\boldsymbol{\varepsilon} + \mathbf{n}_{\textrm{RN}} + \mathbf{n}_{\textrm{CRN}} + \mathbf{n}_{\textrm{WN}} + \mathbf{s}.
\end{equation}

The first term $M\boldsymbol{\varepsilon}$ represents inaccuracies originating from subtracting the linearized timing model solution. Next the term $\mathbf{n}_{\textrm{RN}}$ denotes effects due to low-frequency ``red'' noise that are intrinsic to each pulsar. The following term $\mathbf{n}_{\textrm{CRN}}$ again describes red-noise sources, but this time specifically references sources that are common amongst all of the pulsars, including for example a GWB. Here we model the common spectrum process with a fiducial power-law spectrum with a characteristic strain $h_{c}$ and cross-power spectral density $S_{ab}$:

\begin{align}
    h_{c}(f) &= A_{\mathrm{gw}}\left(\frac{f}{f_{yr}}\right)^{\alpha}, \\
    S_{ab}(f) &= \Gamma_{ab}\frac{A_{\mathrm{gw}}^{2}}{12\pi^{2}} \left(\frac{f}{f_{\mathrm{yr}}}\right)^{-\gamma} f_{\mathrm{yr}}^{-3},
\end{align}

\noindent where the spectral index $\gamma = 3 - 2\alpha$. In the case where the common-process signal represents a background generated by the GW emission from a population of inspiraling SMBHBs in circular orbits, we have $\alpha=-2/3$ and $\gamma=13/3$ \cite{Phinney2001}. The function $\Gamma_{ab}$, called the overlap reduction function (ORF), defines the average correlations between a set of two pulsars $a$ and $b$ based on their relative angular separation. For common uncorrelated red-noise (CURN) processes the ORF is equal to 1. For an isotropic, stochastic GWB it is given by the Hellings-Downs correlation function \cite{1983HellingsDowns}:

\begin{equation}\label{eq:hd}
    \Gamma_{ab} = \frac{3}{2}x_{ab}\ln x_{ab} - \frac{x_{ab}}{4} + \frac{1}{2} + \frac{\delta_{ab}}{2} \,,
\end{equation}

\noindent with $x_{ab} = (1 - \cos\xi_{ab})/2$ for an angular separation $\xi_{ab}$ between two pulsars. Following the common-process noise signals is a term $\mathbf{n}_{\textrm{WN}}$ encoding all high-frequency ``white'' noise sources present in the data, including constant multiplicative correction factors to TOA uncertainties (EFAC), additional noise added in quadrature (EQUAD), and observational epoch-correlated noise (ECORR) that is uncorrelated across separate epochs. Lastly the vector $\mathbf{s}$ represents the component of the timing residuals caused by additional deterministic signals. Here we treat $\mathbf{s}$ as the signal induced by an individual binary, and is outlined in more detail in Sec. \ref{sec:cw}.

Finally, we construct the form of the PTA likelihood for use in our Bayesian inference pipelines. First we dramatically reduce the dimensionality of our posterior by marginalizing over the timing model parameters \cite{2013PhRvD..87j4021L, 2014PhRvD..90j4012V}. Then by constructing the total PTA covariance matrix $C = N + TBT^{T}$, with $N$ the white noise covariance matrix, $T$ the design matrix for the timing model, red noise, and ECORR signals, and $B$ the prior covariance matrix for those three sets of parameters, we can state the multivariate Gaussian likelihood function used for the analyses in this paper:

\begin{equation}\label{eq:likelihood}
    L\left(\delta\mathbf{t} | \boldsymbol{\theta}\right) =  \frac{\exp\left(-\frac{1}{2}\left(\delta\mathbf{t} - \mathbf{s}\right)^{T}C^{-1}\left(\delta\mathbf{t}-\mathbf{s}\right)\right)}{\sqrt{\det 2\pi C}},
\end{equation}

\noindent where $\boldsymbol{\theta}$ denotes the vector of varying parameters in our model.

\subsection{CW Signal}
\label{sec:cw}

We now review the signal model for GWs originating from an SMBHB and their effect on PTA residuals. The GW signal can be written as:

\begin{equation}\label{eq:s_cw}
    s(t, \hat{\Omega}) = F^{+}(\theta, \phi, \psi)\Delta s_{+}(t) + F^{\times}(\theta, \phi, \psi)\Delta s_{\times}(t),
\end{equation}

\noindent where the scripts $\{+ \times\}$ denote the plus and cross polarization modes, respectively, the two tensor polarizations allowed by general relativity, and $\hat{\Omega}$ is a unit vector pointing from the GW source to the solar system barycenter. The functions $F^{+}$ and $F^{\times}$ represent the antenna pattern functions that describe the response of a given pulsar to the emitting source, and are composed of the binary sky location parameters $(\theta, \phi)$ and GW polarization angle $(\psi)$. The terms $\Delta s_{+, \times}(t)$ account for the fact that the Earth and pulsar see the induced GW signal at different times in the binary evolution, and therefore define the difference between the ``pulsar-term'' and ``earth-term'':

\begin{equation}\label{eq:delta_s}
    \Delta s_{+,\times}(t) = s_{+, \times}(t_{p}) - s_{+, \times}(t),
\end{equation}

\noindent where $t_{p}$ is the time measured at the pulsar and $t$ the time measured at the solar system barycenter. For the analyses present in the remainder of the paper, we focus only on searching for the earth-term component of the full signal:

\begin{equation}\label{eq:earthterm}
    s_{E}(t, \hat{\Omega}) = F^{+}(\theta, \phi, \psi)s_{+}(t) + F^{\times}(\theta, \phi, \psi)s_{\times}(t).
\end{equation}

The exact forms of $s_{+, \times}(t)$ for a circular binary are given, to zeroth Post-Newtonian (0-PN) order, by:

\begin{align}
    s_{+}(t) &= - \frac{\mathcal{M}^{5/3}}{d_{L}\omega(t)^{1/3}} \sin 2\Phi(t) \left(1 + \cos^{2}\imath \right), \\
    s_{\times}(t) &= \frac{\mathcal{M}^{5/3}}{d_{L}\omega(t)^{1/3}} 2\cos 2\Phi(t)\cos\imath.
\end{align}

\noindent The parameter $\mathcal{M}$ represents the binary chirp mass $\mathcal{M}\equiv (m_{1}m_{2})^{3/5}/(m_{1} + m_{2})^{1/5}$ for the component black hole masses $m_{1}$ and $m_{2}$. The parameters $d_{L}$ and $\imath$ are the luminosity distance to the binary and the source inclination angle, respectively. The time-dependent angular frequency and phase functions are, for reference Earth-term frequency $\omega_{0}$ and phase $\Phi_{0}$:

\begin{align}
    \omega(t) &= \omega_{0} \left[1 - \frac{256}{5}\mathcal{M}^{5/3}\omega_{0}^{8/3} (t-t_{0})\right]^{-3/8}, \\
    \Phi(t) &= \Phi_{0} + \frac{1}{32}\mathcal{M}^{-5/3} \left[\omega_{0}^{-5/3} - \omega(t)^{-5/3}\right].
\end{align}

Additionally, one can define the overall strain amplitude, $h_{0}$, as

\begin{equation}\label{eq:strain}
    h_{0} = \frac{2\mathcal{M}^{5/3}\left(\pi f_{\mathrm{GW}}\right)^{2/3}}{d_{L}},
\end{equation}

\noindent with the GW frequency $f_{\mathrm{GW}}$ related to the initial angular frequency $\omega_{0}$ by $\omega_{0} = \pi f_{\mathrm{GW}}$. We note that Eq. \ref{eq:strain} shows a degeneracy between $h_{0}$, $\mathcal{M}$, $f_{\mathrm{GW}}$, and $d_{L}$, allowing us to choose three of those four quantities when constructing our complete parameter vector. In practice we typically exclude the luminosity distance and sample $h_{0}$, $\mathcal{M}$, and $f_{\mathrm{GW}}$ alongside the remaining CW source and PTA noise parameters.

\begin{figure}
    \includegraphics[width=\columnwidth]{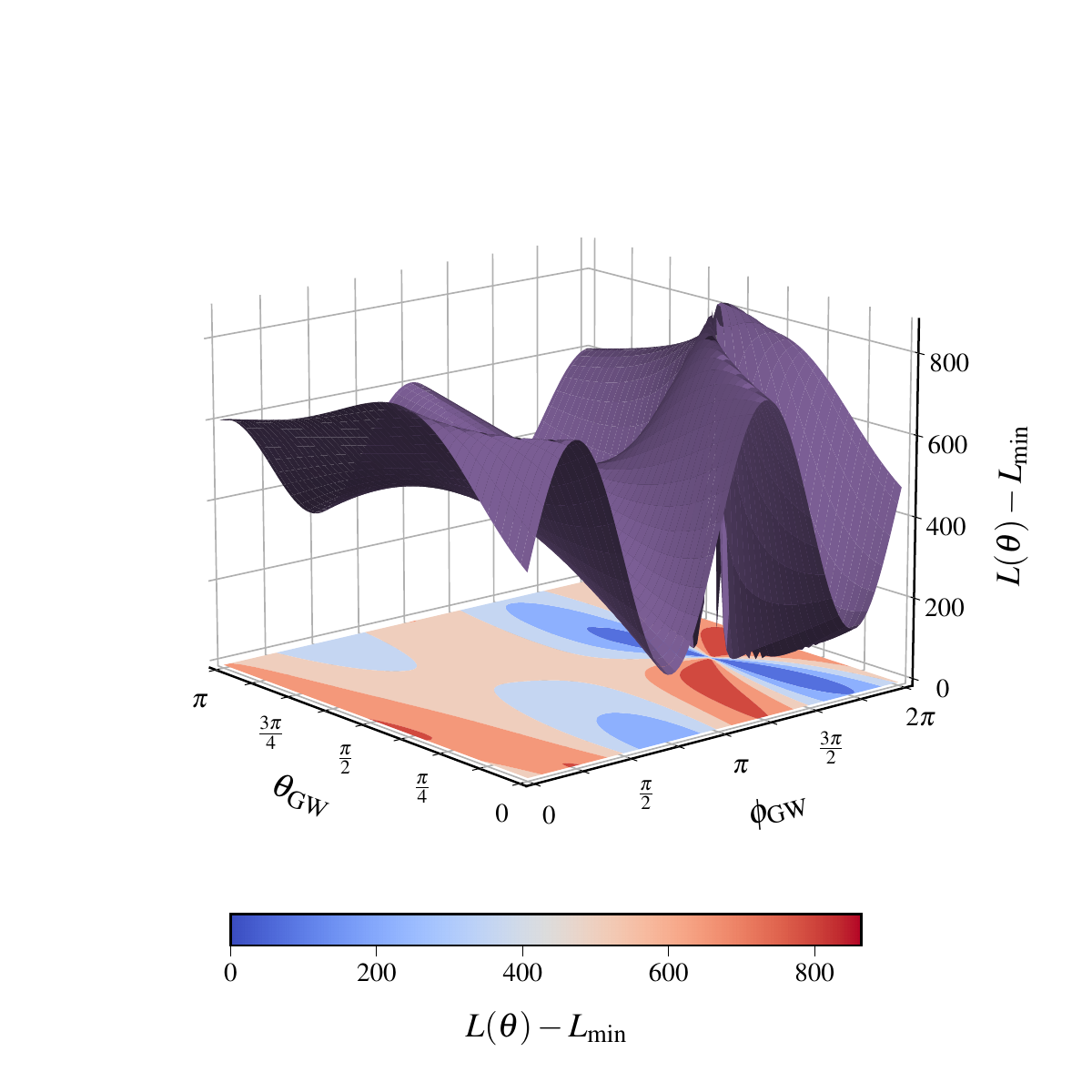}
    \caption{Likelihood surface for an earth-term only CW signal as a function of sky position. The $x$ and $y$ axes represent $\theta_{\textrm{GW}}$ and $\phi_{\textrm{GW}}$ for the source, respectively. The $z$ axis shows the PTA log-likelihood function evaluated at a particular ($\theta_{\textrm{GW}}$, $\phi_{\textrm{GW}}$), then subtracting off the minimum log-likelihood value for the grid. On the plane $z=0$ we plot a 2D colormap contour of the surface. We see that the contours of the likelihood surface have many sharp peaks and valleys, indicating difficult regions of parameter space to sample over.}
    \label{fig:ll_plot}
\end{figure}

In order to further elucidate the complications in sampling joint CW and common-process models, we compute Eq. \ref{eq:likelihood} for an individual binary and particular noise realization. We plot the earth-term only likelihood surface in Fig. \ref{fig:ll_plot} as a function of the CW sky location parameters. The surface displays highly nontrivial structure and demonstrates some of the difficulties in efficiently sampling the full parameter space. There are numerous local extrema where a random-walk MCMC sampler is liable to get trapped and be unable to fully explore the full posterior. This highlights the need for more sophisticated sampling routines and corresponding pipelines.

\subsection{Hamiltonian Monte Carlo Sampling}
\label{sec:hmc}

The HMC algorithm \cite{1987HMCDuane, 2011HMCNeal}, an extension of the traditional Metropolis-Hastings technique \cite{Metropolis-Hastings}, tackles the problem of sampling high-dimensional and covariant state spaces by using Hamiltonian dynamics to generate proposal distributions at distant states, thereby reducing the overall correlation in the Markov chain. It proceeds by first introducing an auxiliary momentum vector $\mathbf{r}$ alongside the model parameters $\boldsymbol{\theta}$. The Hamiltonian to be simulated is the log of the joint density of $\mathbf{r}$ and $\boldsymbol{\theta}$:

\begin{equation}
    H\left(\mathbf{r}, \boldsymbol{\theta}\right) = U\left(\boldsymbol{\theta}\right) + K\left(\mathbf{r}\right) = -\mathcal{L}(\boldsymbol{\theta}) + \frac{1}{2}\mathbf{r}^{T}M^{-1}\mathbf{r},
\end{equation}

\noindent where $\mathcal{L}(\boldsymbol{\theta})$ is the log of the likelihood function for the target parameter distribution, and $M$ a ``mass matrix'' typically taken as the identity. The evolution of this system through time can then be simulated by numerically solving Hamilton's equations:

\begin{equation}
    \frac{d\boldsymbol{\theta}}{dt} = \frac{\partial H}{\partial \mathbf{r}}, \qquad
    \frac{d\mathbf{r}}{dt} = -\frac{\partial H}{\partial \boldsymbol{\theta}}.
\end{equation}

\noindent This integration proceeds for a set number of steps $L$, and ends by proposing some final position and momentum states.

The relative performance of the HMC algorithm is then determined by two factors: the computational cost of calculating the gradient of the log likelihood function for the target distribution, and the proper tuning of two user-defined parameters:  the number of leapfrog steps $L$ and integration step size $\epsilon$. The first factor is driven entirely by the complexity of the model in question, and whether or not the log likelihood gradient can be computed exactly or requires numerical differentiation. The second factor can be resolved by automatically tuning the two extra parameters through the use of the No-U-Turn Sampler \cite{HoffmanGelmanNUTS}, which uses a recursive doubling algorithm that monitors the trajectory of proposal generation and stops when the trajectory makes a ``U-turn,'' or begins to double back on itself.

\subsection{Software}
\label{sec:software}

Our new code, freely and publicly available on GitHub under the package \texttt{etudes} \footnote{\href{https://github.com/gabefreedman/etudes}{https://github.com/gabefreedman/etudes}}, includes an analysis suite capable of performing HMC sampling with PTAs. Although this paper focuses on joint searches for a CW signal and common red-noise process, the modularity of the code allows for the addition of a wide array of other GW sources of interest, such as multiple binaries \cite{2012PhRvD..85d4034B, 2020CQGra..37m5011B}, GW memory \cite{NG12.5yrBWM}, eccentric binaries \cite{2016ApJ...817...70T, 2020PhRvD.101d3022S, 2023CQGra..40o5014S, 2024ApJ...963..144A}, or advanced pulsar noise modeling \cite{2023ApJ...951L...7R, 2024arXiv240514941L}. It can also be natively run on GPUs, drastically dropping the runtime of CW analyses to timescales of hours for simulated data and days for production data. The code is under further active development to accommodate other searches of interest.

While there are renewed efforts towards utilizing hierarchical modeling for PTAs \cite{2024arXiv240605081V}, This work solely uses the marginalized PTA likelihood, meaning that we need not apply coordinate transformations, such as a decentered reparameterization, designed to deal with Markov chain mixing rates and other sampling issues commonly associated with hierarchical funneling. Instead all analyses here, and the default setup for our pipeline, use a single change of coordinates known as an interval transform. This maps all model parameters from their default prior ranges $\theta\in[a, b]$ to the real line $\theta' \in(-\infty, \infty)$ via:

\begin{align}
    \theta' &= \log\left(\frac{\theta - a}{b - \theta}\right), \\
    \theta &= \frac{\left(b-a\right)\exp\left(\theta'\right)}{1+\exp\left(\theta'\right)},
\end{align}

\noindent where we use the Jacobian $d\theta'/d\theta$ and its reciprocal to convert between the original and transformed probability spaces.

Sampling with HMC necessarily requires taking derivatives of the model likelihood. We accomplish this by writing the PTA likelihood and its components computations entirely with \texttt{JAX} \cite{jax2018github}, allowing us to use automatic differentiation to calculate gradients. To do so we decouple the the entirety of the PTA computation from NANOGrav's analysis suite \texttt{enterprise} \cite{enterprise}, though we do make use of the code's data structures for holding per-pulsar TOAs, residuals, and other timing model information. We utilize the implementation of the NUTS algorithm present in the \texttt{blackjax} \cite{cabezas2024blackjax} package. All simulated PTA datasets are created using \texttt{libstempo} \cite{libstempo}.

\section{Simulated Data Study}
\label{sec:sim_data}

First in order to gauge the accuracy of our pipeline and demonstrate its consistency in parameter estimation we created and analyzed a collection of simulated datasets. All datasets comprise identical TOAs, uncertainties, and timing model solutions to the NANOGrav 12.5-year dataset \cite{NG12.5Dataset}. This constitutes 45 pulsars in total, all with an observational baseline of at least 3 years.

Each individual dataset contains the same per-pulsar noise injections. We simulated ``white-noise'' signals, typically instrumental noise that dominates at high frequencies, at their maximum likelihood values obtained from separate individual pulsar noise analyses. Low-frequency ``red-noise'' signals, representing noise intrinsic to each pulsar, were simulated again by referencing the same individual noise runs. We injected the intrinsic pulsar noise at frequencies spanning from $1/T_{\textrm{psr}}$ up to $30/T_{\textrm{psr}}$, with $T_{\textrm{psr}}$ denoting the observational timespan of each pulsar.

The NANOGrav 12.5-year dataset contained a CURN process with a Bayes factor in excess of 10,000 relative to a model with only intrinsic pulsar noise \cite{NG12.5GWB}. Therefore, for the most accurate prescription of a realistic PTA dataset, we also include a similar process in all of our simulations. The most recent dataset reported evidence for this process containing HD correlations, though we do not consider that in this study. We inject a CURN signal characterized by an amplitude $A_{\textrm{CURN}} = 2 \times 10^{-15}$ and spectral index $\gamma_{\textrm{CURN}} = 4.33$, in line with the expected power and shape of the spectrum.

On top of the various noise models, we inject CW signals. We choose three instances with which to create our data: a low-frequency source, a high-frequency source, and a dataset with no source injection. In all cases we inject only the earth-term signal. The low-frequency dataset contains a binary emitting GWs at frequency $f_{\textrm{GW}} = 6$ nHz and an amplitude chosen to achieve a moderately high signal-to-noise ratio (SNR) of 10.8. For the case of the high-frequency dataset, we include a binary emitting GWs at $f_{\textrm{GW}} = 60$ nHz with an SNR of 9.3. In both cases the SNR is calculated as:

\begin{equation}\label{eq:snr}
    \mathrm{SNR} = \sqrt{\left(s | s\right)} = \sqrt{s^{T}C^{-1}s},
\end{equation}

\noindent where $s$ is the template waveform and. This can also be considered the expected SNR that is independent from any particular noise realization. The dataset without any CW injection allows us to verify the ability of our methods to place upper limits on source properties in the absence of a detection. For the purposes of validating our pipeline, we create 100 simulated datasets with both the 6 nHz and 60 nHz injection properties. This allows us to test our methods across numerous noise realizations.

Next we outline the basic procedure for setting up our models before performing Bayesian inference through our HMC pipeline. Rather than simultaneously search over the hundreds of white-noise parameters, we fix them to their maximum likelihood values used in creating the datasets, a commonplace procedure in production-level PTA analyses. We model the pulsar intrinsic red-noise with a power-law power spectral density (PSD) defined by an amplitude $\log_{10}\, A_\mathrm{red} \in U[-18, -11]$ and spectral index $\gamma_{\mathrm{red}} \in U[0, 7]$. Additionally we search over the two parameters characterizing the CURN process, using priors of $A_\mathrm{CURN} \in U[-18, -12]$ and $\gamma_\mathrm{CURN} \in U[0, 7]$. When modeling the CW signal all corresponding parameters are given uniform priors. In the case of upper limit analyses, the prior on $\log_{10}h_{0}$ is shifted from uniform in log space $\log_{10}h_{0} \in U[-18, -12]$ to uniform in linear space $\log_{10}h_{0} \in U[10^{-18}, 10^{-12}]$. The coordinate space outlined by these priors is later transformed via the procedure outlined in Sec. \ref{sec:software} prior to beginning the inference.

Lastly we benchmark the speed and efficiency of both the \texttt{etudes} pipeline and comparable run with \texttt{enterprise} through a pilot inference run on one of the simulated 6 nHz injection datasets. The full joint CURN and CW search here constitutes 100 free parameters ($2N_{\textrm{psr}}$ intrinsic red-noise parameters for 45 pulsars, 2 parameters for the CURN, and 8 describing the CW signal model). For the traditional MCMC pipeline with \texttt{enterprise} the average likelihood evaluation time is 200 ms on a 12-core Intel(R) Xeon(R) E5-2680 v3 processor. Using the same CPU, the average likelihood and gradient evaluation times with \texttt{etudes} is 170 ms and 3.7 s, respectively, and on an NVIDIA Tesla A100 GPU they are 17 ms and 1.6 s, respectively. All of the above results are then scaled by the average autocorrelation lengths of the corresponding MC chains to calculate the timescales of statistically independent sample generation. This gives an estimate of 450 s to get an independent sample with \texttt{enterprise} compared to 52 s for runs on a CPU and 25 s on a GPU for the HMC pipeline. Overall, Hamiltonian sampling provides an increase of roughly an order of magnitude in computational efficiency. 

\subsection{Low-frequency (6 nHz) Signal}

Previous NANOGrav CW searches have consistently shown that PTAs are most sensitive to single sources at the lower end ($\sim1-20$ nHz) of their frequency ranges. This also happens to be where the GWB, and more generally any CURN process, is at its strongest. With evidence for a GWB now in hand, it is important for all future CW searches to be capable of dealing with the covariance between the common signal and any low-frequency single sources. We first analyzed a signal with a frequency of $f_{\textrm{GW}} = 6$ nHz, which places it near the peak sensitivity of NANOGrav PTA. The chirp mass $\mathcal{M} = 10^{9} M_{\odot}$ and luminosity distance $d_{L} = 21.8$ Mpc of the source are chosen so that the GW amplitude gives an SNR of 10.8. Additionally we place the source close to the most sensitive sky location at $(\theta, \phi) = (2\pi/3, 3\pi/2)$. Lastly, the parameters $(\imath, \psi, \Phi_{0}) = (3\pi/4, \pi/3, 3\pi/2)$ define the source's inclination, polarization, and initial phase. Together, all of the above allows to fully classify our injected signal.

\begin{figure*}[t]
    \includegraphics[width=2\columnwidth]{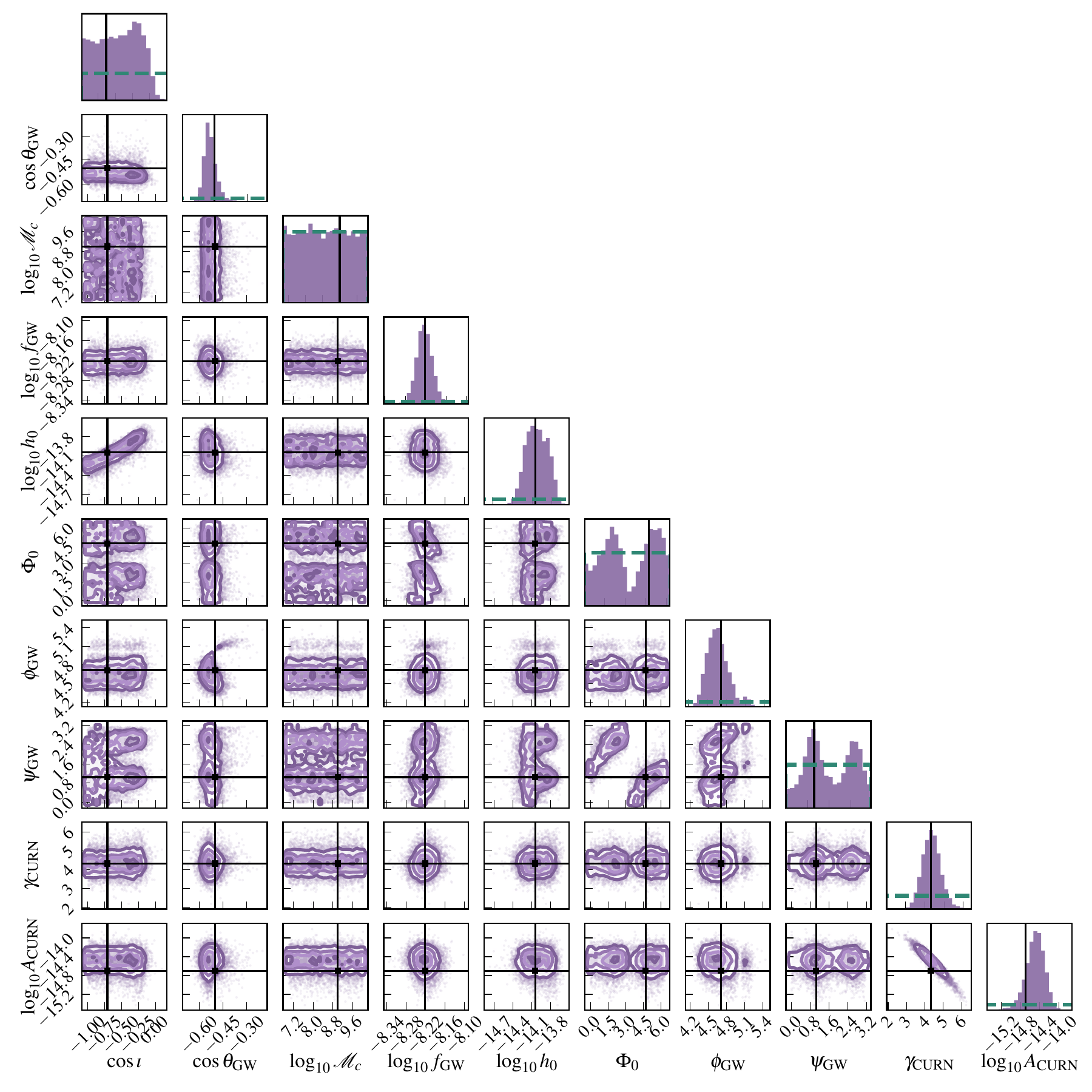}
    \caption{1D and 2D posterior distributions for the eight parameters describing a SMBHB signal emitting GWs at $f_{\textrm{GW}} = 6$ nHz at an SNR of 10.8. The true values of the injected parameters are shown as solid black lines, and the priors are plotted on the 1D histograms as horizontal, green dashed lines. All true values fall within their posteriors, with the sky location, GW frequency, and GW strain parameters being tightly constrained. This demonstrates the capability of the HMC pipeline in accurate parameter estimation for full CW searches.}
    \label{fig:lowfreq_sim}
\end{figure*}

Taking the resulting chains from our analyses, we plot both the one- and two-dimensional posterior distributions for all eight binary parameters in Fig. \ref{fig:lowfreq_sim}. All parameters have their true injected values lying within their respective posteriors. Both the GW frequency and amplitude distributions are tightly constrained. The posterior for the chirp mass remains entirely unconstrained as we expect for earth-term only searches and sources with slow frequency evolution. The sky location of the source is very well localized to its true value. The initial phase and polarization angles display a set of multimodal posteriors, which we can efficiently sample but are unable to break the multimodality.

\subsection{High-frequency (60 nHz) Signal}

The long-term prospects of CW detection play crucial role in multi-messenger analyses and astrophysical interpretation of SMBHB populations and sources, and it is important that we have the ability to do accurate parameter estimation on possible binary candidates. With this in mind, we analyzed a signal with a GW frequency of $f_{\textrm{GW}}=60$ nHz, chosen to closely mimic that of the potential SMBHB candidate 3C 66B \cite{2004ApJ...606..799J, 2010ApJ...724L.166I, 2024ApJ...963..144A}. The remaining parameters describing the source properties and sky location are $(\theta, \phi, \imath, \psi, \Phi_{0}, \mathcal{M}_{c}, d_{L}) = (2\pi/3, 3\pi/2, 3\pi/4, \pi/3, 3\pi/2, 10^{9} M_{\odot}, 91.1 \textrm{Mpc})$. At frequencies this high the CURN is very weak and therefore we do not have to worry with covariances between the common-process and binary signals. 

\begin{figure*}[t]
    \includegraphics[width=2\columnwidth]{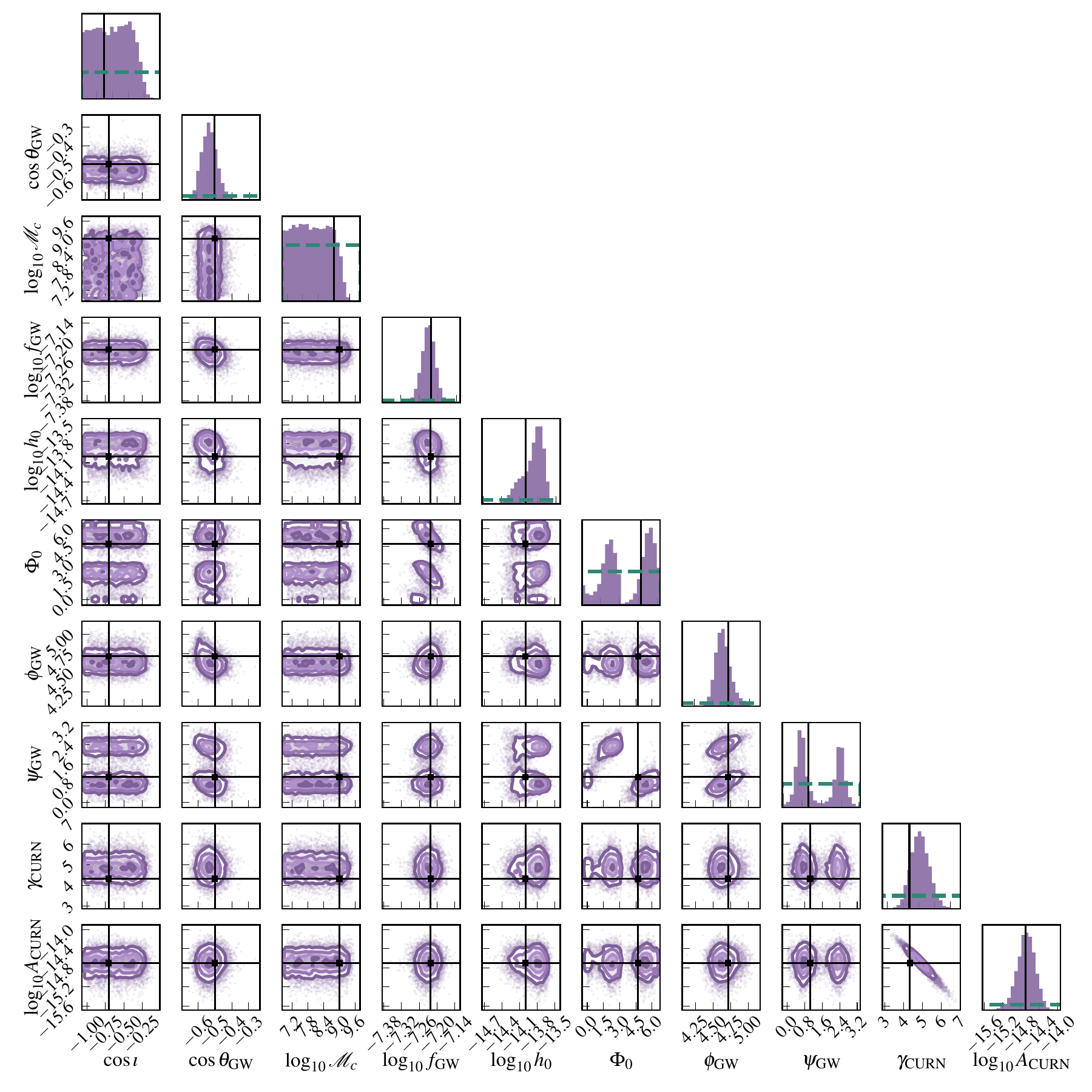}
    \caption{1D and 2D posterior distributions for the eight parameters describing a SMBHB signal emitting GWs at $f_{\textrm{GW}} = 60$ nHz at an SNR of 9.3. The true values of the injected parameters are shown as solid black lines, and the priors are plotted on the 1D histograms as horizontal, green dashed lines. Similar to the low-frequency injection analysis, all true values fall within their posteriors, with parameters such as the sky location, GW frequency, and GW strain parameters being tightly constrained. The binary chirp mass posterior now features an upper limit excluding sources that would have undergone significant frequency evolution over the data timespan.}
    \label{fig:highfreq_sim}
\end{figure*}

In Fig. \ref{fig:highfreq_sim} we plot the posterior distributions for the eight binary parameters for this model, similar to Fig. \ref{fig:lowfreq_sim}. Again we find that we are able to efficiently sample the entire CW parameter space alongside both a CURN process as well as all intrinsic pulsar noise. We see the same structure in the nearly all of our posteriors: the source GW frequency, GW amplitude, and sky location are very tightly constrained, and the multimodal structure in the polarization angle and initial phase persist. Most importantly, all injected values once again fall squarely within their 1D posteriors. One notable difference is the emerging constraint on the binary chirp mass. High-mass binaries emitting at this frequency should show significant evolution over the 12.5-year observing window of our simulated datasets. Consequently we find across the 100 realizations of the data that we can place an upper limit on the binary chirp mass. In 30 of the realizations, the chirp mass posterior was less constrained than what the frequency evolution would predict.

\subsection{Parameter Estimation Consistency}

As a final test of our method's effectiveness with simulated data, we explore the capacity of its statistical coverage across many noise realizations of the same underlying data. First we create 100 iterations of our $f_{\textrm{GW}}=6$ nHz dataset. Next we run standard Bayesian searches on all datasets with our HMC pipeline. Lastly, to check the consistency of parameter recovery for our pipeline, we consider across all 100 sets of posteriors whether if in $p$\% of the realizations the injected parameter values fall within the $p$\% credible region.

\begin{figure}[t]
    \includegraphics[width=\columnwidth]{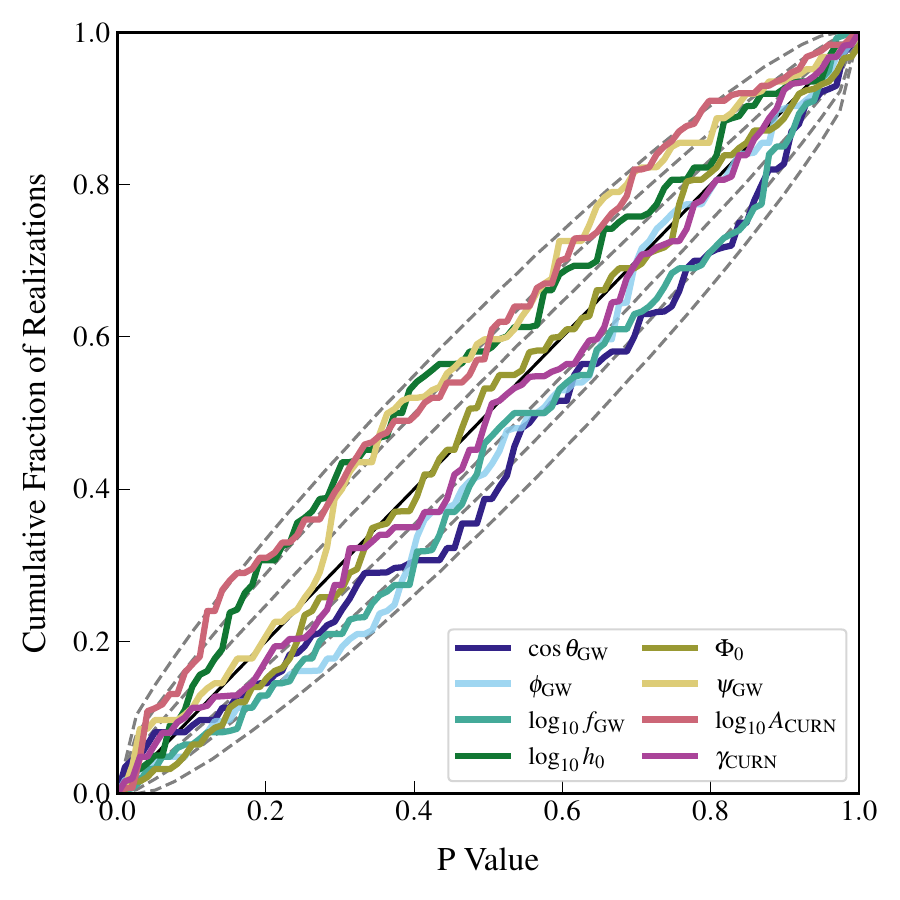}
    \caption{$p-p$ plot displaying recovery of injected parameters across 100 simulated PTA datasets. All datasets contain a CURN process and a 6nHz CW injection. Plotted are six lines corresponding to the CW sky location parameters, log strain, log frequency, and CURN amplitude and spectral index. The solid black line along the diagonal represents the line of perfect recovery. Dotted gray lines represent $1\sigma$, $2\sigma$, and $3\sigma$ confidence intervals. All plotted parameters lie within these boundaries indicating no significant bias in parameter recovery.}
    \label{fig:pp_plot}
\end{figure}

The results of this analysis, called a $p-p$ plot, are summarized in Fig. \ref{fig:pp_plot}. We plot lines for CW sky location parameters, $\log_{10}h_{0}$, $\log_{10}f_{\textrm{GW}}$, and the CURN amplitude and spectral index. The dotted gray lines represent $1\sigma$, $2\sigma$, and $3\sigma$ confidence intervals. All parameters fall largely within the $3\sigma$ boundary indicating an unbiased recovery of the injected values. The chirp mass, being entirely unconstrained across all realizations due to the minimal evolution of the particular signal, was left out off this figure. The cosine of the binary inclination was also largely unconstrained across all realizations and was likewise excluded.

\section{Analysis of Real PTA Data}
\label{sec:real_data}

Ultimately we want to validate our methods against real data and published results. We use the full NANOGrav 12.5-year dataset \cite{NG12.5Dataset} to benchmark our analysis, and focus on the particular challenge of creating sensitivity sky maps. Given the anisotropic distribution across the sky of the pulsars in our array, it is important to quantify how our observing limits change in different areas. The sky maps typically describe, for a given GW frequency, the 95\% upper limits on $h_{0}$ as a function of sky location. The typical strategy for generating the plots is to bin the sky into 768 separate pixels and run an MC analysis on each individual partition. This dense pixelation is due in part to our inability to get similar number of MCMC samples across the full parameter space in an all-sky search. 

We analyze a CW model including a CURN process for sky locations bounded by $\theta\in\left[\pi/2, 3\pi/4\right]$, $\phi\in\left[3\pi/2, 2\pi\right]$. The bounds were chosen so as to include the most sensitive sky location from the NANOGrav 12.5-year CW analysis \cite{NG12.5CW}, at an RA of $19^{\textrm{h}}07^{\textrm{m}}30^{\textrm{s}}$ and a Dec of $30\degree00'00''$. This range of parameter space corresponds to 72 distinct pixels, and therefore typically 72 independent analyses, in the resolution of the sky map from the NANOGrav 12.5-year CW paper. The CW frequency is held fixed at $f_{\textrm{GW}}=7.65\times10^{9}$ Hz, the most sensitive frequency in the NANOGrav 12.5-year dataset.

Our strategy is to run one single chain with HMC sampling and leverage the pipeline's efficiency to fully explore across the broader sky range, allowing us to compute a series of GW strain upper limits as a function of sky location and populate the sky map in post-processing. We run one single analysis for $M=80,000$ samples, after which we break up our chains into sky location bins consistent with the full 768-pixel map. With all autocorrelation lengths of order $\mathcal{O}(1)$, after thinning this results in between $700-1,200$ independent samples per reduced sky pixel.

\begin{figure}
    \includegraphics[width=\columnwidth]{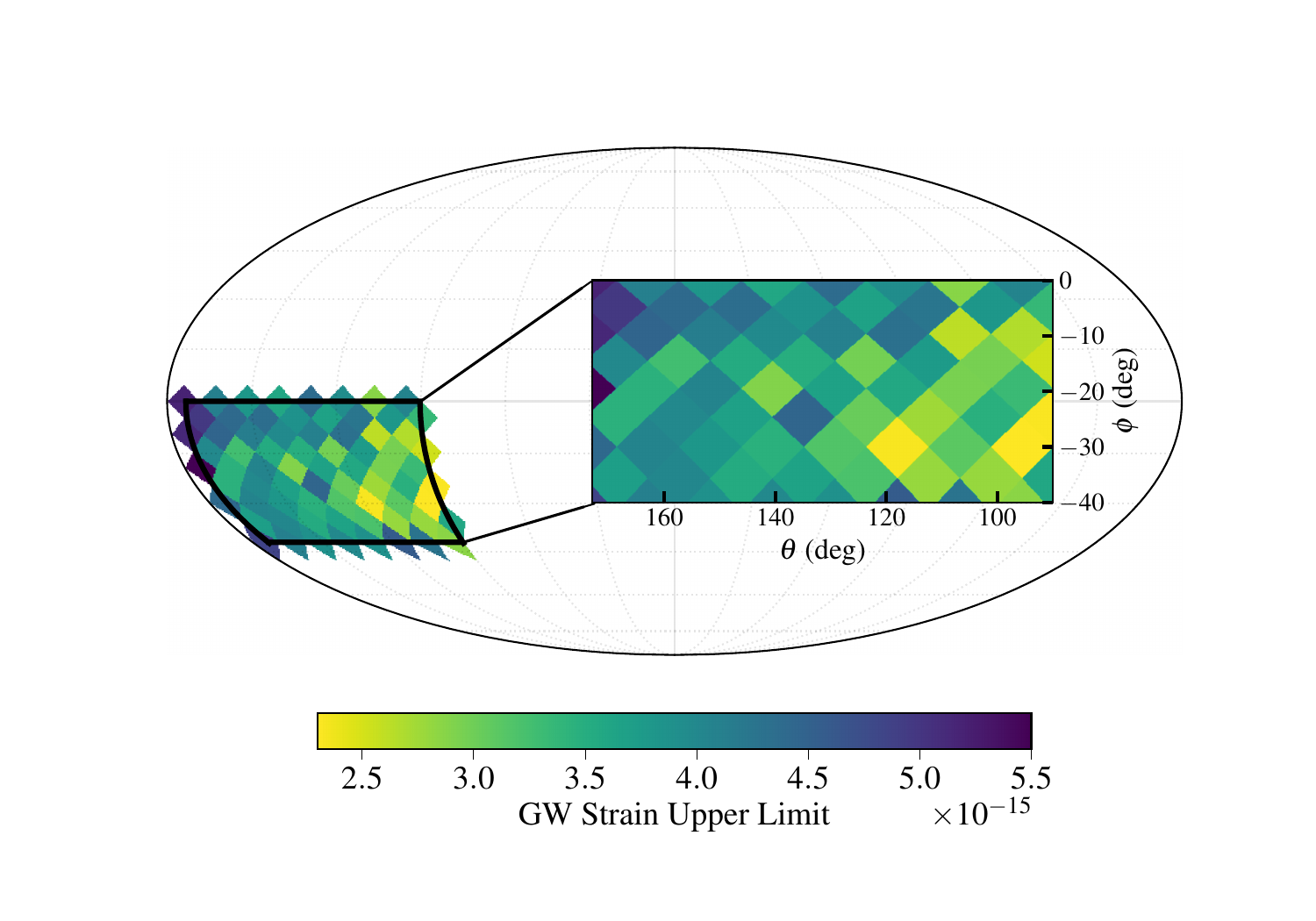}
    \caption{Map displaying CW strain 95\% upper limits for a range of sky location parameters bounded by $\theta\in\left[\pi/2, 3\pi/4\right]$, $\phi\in\left[3\pi/2, 2\pi\right]$. The data are taken from a single chain run with an HMC pipeline and pixelated to match the resolution of the analogous map for the 12.5-year data set. The analysis is run for $f_{\textrm{GW}}=7.65\times10^{9}$ Hz, the most sensitive frequency searched. Pixel to pixel uncertainties range between $1.03\times10^{-16} < \sigma_{h_{0}} < 1.81\times10^{-15}$.}
    \label{fig:skymap}
\end{figure}

In Fig.~\ref{fig:skymap}, we plot the results of our reconstructed sky map. The strain upper limit at the most sensitive sky location is $h_{0} < (2.15 \pm 0.30)\times 10^{-15}$. Its coordinates exactly match that of the most sensitive region from the full NANOGrav 12.5-year analysis, which reported a strain upper limit for that pixel of $h_{0} < (2.66 \pm 0.15)\times 10^{-15}$ \cite{NG12.5CW}. We find the upper limit at the least sensitive sky location to be $h_{0} < (5.45 \pm 0.36)\times 10^{-15}$. Unlike the corresponding analysis and map in the NANOGrav 12.5-year CW paper, we marginalize over the amplitude and spectral index of the CURN process instead of fixing the signal parameters to their maximum likelihood values. We also only search over the Earth term of our CW signal. Therefore we do not expect to find perfect agreement between the two results when comparing on a pixel-to-pixel basis.

By effectively sampling over larger portions of the sky, we can cut the computational cost of generating a full sky map by nearly an order of magnitude. Increasing the pixel range of our searches is also a step closer to eliminating a grid-based structure in the otherwise fully Bayesian analysis. The limiting factor in expanding the prior range is purely the computational wall time rather than specific choices on location binning as the HMC sampler can fully explore the posterior even at the least-preferred sky locations. With enough time this can develop into a single all-sky search for producing upper limit maps, and more easily enable making the maps at many different GW frequencies of interest.

\section{Discussion}
\label{sec:conclusions}

In this paper we have presented an end-to-end pipeline for performing efficient Bayesian searches of the high dimensional and complicated parameter spaces for joint CW and common red-noise process signal analyses with PTA data. Our code employs HMC sampling to conduct accurate parameter estimation. We demonstrated the performance of this sampling routine through numerous tests across both simulated and real PTA data. We showed that by using HMC sampling we can effectively do parameter estimation for both high- and low-frequency CW signals. The methods are robust towards conducting these analyses while simultaneously marginalizing over a common-process signal and can accurately recover both GW signals.

By utilizing the HMC algorithm as our default underlying sampler, we are able to both significantly lower the autocorrelations in our MCMC chains as well as reduce the total number of samples we require per run. Our ability to evenly sample wider areas of the sky means that we are closer to removing a binning element of our otherwise completely Bayesian analysis. The sampler also scales favorably with dimensionality, a positive sign as future PTA datasets inch closer to containing $\mathcal{O}(100)$ pulsars and $100$s of corresponding noise parameters.

A significant long-term advantage of this pipeline is its modularity and ability to adapt to a wide range of signal modeling choices. The code is not designed solely for the task of CW searches and can develop and grow into a general purpose analysis suite similar to the current analysis suite \texttt{enterprise}. For example, it can modified to run on models considering only a GWB signal, for which previous efforts have already shown HMC sampling to be increasingly useful \cite{HMCandPTAs_Freedman}. Further development can also add the possibility of more sophisticated pulsar noise models or additional deterministic sources of interest in the PTA band. The future of PTA GW analyses is in part defined by its potential computational pitfalls: an ever-increasing data span, noise modeling of growing complexity, and the goal of combined international datasets. These methods will prove a valuable tool alongside the range of computational developments in the PTA community towards addressing these issues before they arise, and keeping our analyses tractable to the future.

\acknowledgments
We thank Shashwat Sardesai for useful discussions throughout this project. We also thank Bence B\'ecsy for valuable and insightful comments on this manuscript. G.E.F. is supported by National Aeronautics and Space Administration (NASA) Future Investigators in NASA Earth and Space Science and Technology Grant No. 80NSSC22K1591. The authors are members of the North American Nanohertz Observatory of Gravitational Waves (NANOGrav) collaboration, which receives support from NSF Physics Frontiers Center Award No. 2020265.

\bibliographystyle{apsrev}
\bibliography{authors}

\end{document}